\documentclass{iopconfser}
\usepackage{graphicx}
\usepackage{doi}

\begin{document}

\noindent
CERN-TH-2024-227
\\

\title{UHECR measurements and physics at
man-made accelerators: mutual constraints}

\author{Maria Vittoria Garzelli$^{\,1,2}$}

\affil{$^1$II Institut fuer Theoretische Physik, Universitaet Hamburg, Luruper 
Chaussee 149, D-22761 Hamburg, Germany}
\affil{$^2$ CERN, Department of Theoretical Physics, CH–1211 Geneva 23, Switzerland}

\email{maria.vittoria.garzelli@desy.de}

\begin{abstract}
Measurements of the products of UHECR interactions with the Earth's atmosphere, as obtained in Extended Air Shower experiments, 
offer important information con\-cer\-ning hadronic interactions,
which for some aspects overlaps and for many others complements the information extracted by measurements of collisions at human-made accelerators. In this contribution I discuss some of the constraints that the UHE astroparticle and accelerator fields exercise one over each other, emphasizing the importance of further new measurements, through new experiments or observations, in both fields.
\end{abstract}

\section{Introduction}
The cosmic ray (CR) spectrum extends on 
scales of many orders of magnitudes, spanning $\sim$ 14 orders in energy and more than 30 orders in flux intensity. Its interpretation requires therefore knowledge of physics in many different conditions. With increasingly more precise measurements along the years, new features have progressively emerged (from first and second knee, ankle, cut-off, to the recently observed instep at $E \sim 1.4 \cdot 10^{19}$ eV~\cite{PierreAuger:2020kuy}). One of the objectives of CR-oriented research programs is understanding the origin of these features, which may emerge due to an interplay of source characteristics, mechanisms of acceleration and properties of particle interactions and propagation from the sources to the Earth. These aspects have been investigated through the observation of CRs conducted by either ground-based experiments at Earth or balloons/satellites, depending on the primary energy, but a lot of questions are still open. 

\section{CR origin, UHE CR observables and the muon puzzle} 
In order to understand CR origin, we are interested in CR primary energy, arrival direction, mass number $A$, event-by-event. A multimessenger approach, considering the simultaneous observation of photon, neutrino and gravitational wave signals, by means of instruments acting in coordination, can be of great help in this respect~\cite{Adhikari:2022sve}.
While direct CR detection is possible for $E < 100$~TeV, only indirect detection has  been achieved so far for $E > 100$~TeV, implying that, at high enough energies, $E$, $A$ and direction have been reconstructed from the products of Extender Air Showers (EAS) generated by the interaction of CRs with the atmosphere. In particular, the direction is reconstructed from the arrival times of EAS particles on the EAS detectors, distributed over very large areas. 
On the other hand, $E$ is inferred from the size of the electromagnetic component, whereas $A$ can be inferred from various observables. The most popular/studied ones are $\langle X_{max} \rangle$ and $N_\mu$. The first one corresponds to the average atmospheric depth of maximum development of the EAS electromagnetic component in the atmosphere, whereas the second one is the number of muons at Earth, a footprint of the EAS hadronic component. 
The CR composition is one of the most pressing problems. The CRs are (charged) ions. Although the showers produced by $\gamma$ rays with energies
from tens of GeV to $\mathcal{O}$(100 TeV)  are commonly observed through Air Cherenkov telescopes, no high-energy $\gamma$ ($E >$ 100 TeV) or $\nu$-initiated shower has been observed so far. While at low energies the ion species can be directly distinguished (with results not always easy to interpret), at high energies one can infer composition from the properties of EAS generated by the CRs interacting with the atmosphere. Considering that none of the hadronic interaction models presently available are able to explain simultaneously all EAS-related observables, big uncertainties affect the mass composition of ultra-high-energy (UHE) CRs. 
In particular, $N_\mu$ predictions computed using as input the composition inferred from $\langle X_{max} \rangle$ observations at each primary energy $E$ are inconsistent with $N_\mu$ data.
One could question how robust are the measurements of these quantities. 
At the Pierre Auger Observatory  measures of  $\langle X_{max} \rangle$ and
$\sigma(X_{max})$ with surface detectors (SD) have allowed to extend the $E$ coverage towards larger $E$, with respect to the fluorescence detectors (FD). Although some systematics in the difference of the results of SD and FD as for $\sigma(X_{max})$ still needs to be understood~\cite{PierreAuger:2023bfx, PierreAuger:2023kjt, PierreAuger:2023gmj}, the most recent measurements conducted with different techniques point to similar overall features concerning CR composition. In particular, the composition, made by mostly protons at $E \sim 3 \cdot 10^{18}$~eV, becomes increasingly heavier at larger energies, but the contribution of elements as heavier as Fe remains small. On the other hand,  (FD+SD) combined data on
 $N_{\mu}$ correspond to 
a systematically heavier composition~\cite{Salamida:2023fmk}. 
This is called ``muon puzzle'' and was already observed in earlier measurements (see e.g.~\cite{PierreAuger:2014ucz, PierreAuger:2016nfk, PierreAuger:2020gxz}).

One interesting point, which increases the difficulty of interpreting UHE CR data, is the fact that CR energies reach values much larger than achievable at the
Large Hadron Collider (LHC), the  most energetic among human-made accelerators. In fact, the highest primary energies recorded for CRs interacting with the Earth's atmosphere correspond to collisions with center-of-mass energies $\sqrt{S} \sim$~300~TeV, i.e. not only well above  LHC possibilities, but even beyond the projections for the Future Circular Collider (FCC), in the modality of hadron-hadron interactions. 
While new physics, still unobserved at the LHC and not yet implemented in the generators of hadronic interactions traditionally used in EAS analyses, could show up at such high energies, making more complicated the 
EAS evolution and description, it has been observed that the muon (and composition) puzzle starts to appear at energies much lower, around $\sqrt{S_{NN}} >$~8~TeV and that the discrepancy between $N_\mu$ predictions and observations is rising gradually with increasing energy~\cite{Albrecht:2021cxw}. This looks more compatible with some shortcoming in the implementation of Standard Model (SM) physics in the
generators for hadronic interactions currently used in EAS simulations, than with the abrupt onset of New Physics effects above a given threshold (e.g. through the production of new heavy beyond-the-SM particles). At the LHC, so far, no sign of New Physics has been found up to $\sqrt{S_{pp}} \le 13.6$~TeV. 

The muon puzzle appears for all generators and for most of the UHE experiments. 
There are few exceptions. One was the measurements of $N_\mu$ at the Haverah Park Array, where the $N_\mu$ computed from the composition inferred from $\langle X_{max} \rangle$ has turned out to be in better agreement with the $N_\mu$ direct measurement than at the Pierre Auger Observatory~\cite{Cazon:2023ojj}. However the case deserves more studies, considering the uncertainties in the absolute energy scale of the water-Cherenkov detectors at Haverah Park, derived from theoretical models.

The generators for hadronic interactions have been investigated in depth, reasoning on possible mo\-di\-fi\-cations within the SM, which would allow to increase the predicted $N_\mu$ recoinciling it with the experimental observations, without spoiling, at the same time, the data/theory agreement for the $\langle X_{max} \rangle$, $\sigma(X_{max})$ and $\sigma(N_\mu)$ observables. Out of four possible modifications, affecting respectively the multiplicity of produced charged particles, the $p$-Air total inelastic cross section, the inelasticity of the interactions, and the produced $\pi^0$ fraction $f_{\pi^0} = N (\pi^0)/N(\pi^0+\pi^++\pi^-)$, only the reduction of $f_{\pi^0}$ turned out to allow to satisfy the previously mentioned criterium~\cite{Albrecht:2021cxw}. 
A reduction of $f_{\pi^0}$ corresponds to a $N_\mu$ increase. In the Heitler-Matthews model, $N_{\mu}$ is proportional to the hadronic energy, which, in turn, is proportional to $(1-f_{\pi^0})^N$, where $N$ is the number of steps in the EAS evolution.  In the case of perfect isospin symmetry (that is what is usually assumed), $f_{\pi^0}=1/3$. Various mechanisms can effectively reduce $f_{\pi^0}$. 

\section{Mechanisms to reduce $f_{\pi^0}$} 

A first mechanism to reduce $f_{\pi^0}$ is breaking isospin symmetry (a breaking is justified because mesons are massive) by $\rho^0$ enhancement with respect to $\pi^0$, followed by $\rho^0 \rightarrow \pi^+ \pi^-$ decay, leading to $N_{\pi^0} /(N_{\pi^+} + N_{\pi^-}) < 1/2$.

\begin{figure}[t!]
\begin{center}
\includegraphics[width=0.70\textwidth]{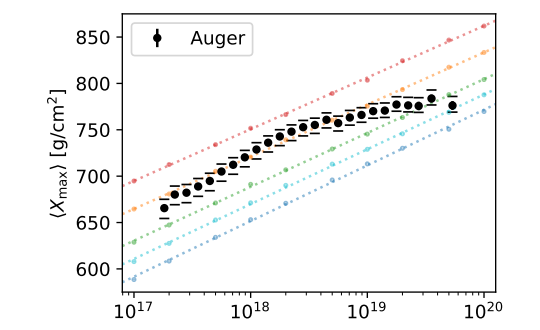}
\includegraphics[width=0.66\textwidth]{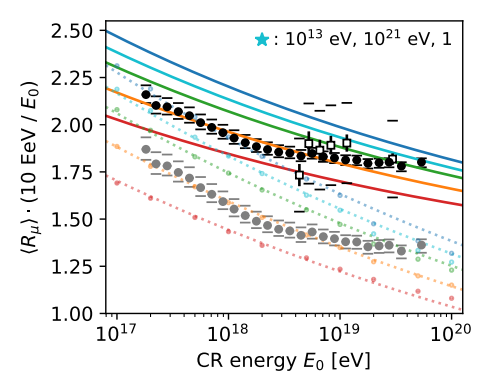}
\end{center}
\caption{\label{fig:strangeb} Composition inference from Pierre Auger data (error bars) on $\langle X_{max} \rangle$ (top) and $\langle R_\mu \rangle$ (bottom)~\cite{PierreAuger:2019phh}, where $R_\mu$ is the total number of muons at ground relative to the total number of muons in a shower with primary energy $10^{19}$ eV, using \texttt{EPOS-LHC} and the strangeball model. 
The application of the strangeball model (solid lines) leaves $\langle X_{max} \rangle$ unaffected. The $\langle X_{max} \rangle$  data (black error bars) can thus be interpreted within the Standard Model (dotted lines). 
On the other hand the strangeball model modifies $\langle R_\mu \rangle$.
In the bottom panel a direct comparison with $\langle R_\mu \rangle$ data 
(white square error bars) follows from mapping $\langle X_{max} \rangle$ data to $R_\mu$ values within the Standard Model (gray error bars) and considering a
strangeball scenario (black error bars) characterized by $E_{min} = 10^{13}$~eV, $E_{max} = 10^{21}$ eV and $n = 1$. Strangeballs are produced for energies in the interval [$E_{min}$, $E_{max}$], $n$ is a parameter regulating the strangeball production probability as a function of the energy. 
The color of the lines refer to the composition of primary CRs: proton (red), helium (orange), nitrogen (green), silicon (cyan), and iron (blue). 
See Ref.~\cite{Manshanden:2022hgf} for more detail.}
\end{figure}

Another one is enhancing strangeness, by increasing the number of produced kaons and/or strange baryons with respect to non-strange hadrons. The possibility of formation of a Quark Gluon Plasma (QGP) phase in CR collisions with the air, or even fireballs~\cite{Anchordoqui:2016oxy}, characterized by an extreme baryochemical potential and extreme temperature, have been proposed. Alternative proposal is the formation of strangeballs~\cite{Manshanden:2022hgf}, which involves strangeness enhancement without the formation of a QGP. An example of the capability of the strangeball model to reproduce data on both $\langle X_{max} \rangle$ and $N_\mu$ is given in Fig.~\ref{fig:strangeb}
In collider physics, strangeness enhancement with respect to predictions computed on the basis of standard shower Monte Carlo (SMC) event generators, such as \texttt{PYTHIA}, has been observed by the ALICE experiment, in $pp$, $p$Pb and PbPb collisions, looking at a variety of produced mesons and baryons ($K_S^0$, $\Lambda^0 + \bar{\Lambda}^0$, $\Xi^\pm$, $\Omega^\pm$) at different $\sqrt{S_{NN}}$~\cite{ALICE:2016fzo}. Considering that this phenomenon affects even small systems, where correlations leading to the formation of ridge and flow have also been observed, the hypothesis has been formulated that a QGP phase might occur in the early stage of the collisions even in the case of small systems. 
This is particularly interesting considering the typical colliding processes leading to EAS formation. 
In this case, local temperature
fluctuations can be large and QGP
droplets with radius depending on the
temperature, instead of a unique
deconfined system of quarks and gluons,
could be formed~\cite{Sahoo:2019ifs}. A somehow connected practical simplified geometrical realization of a similar scenario would be the core-corona model, i.e. the possibility that in the overlapping region of the colliding hadrons/nuclei a ``core'' is formed, 
corresponding to a QGP phase, whereas the region where the colliding systems do not overlap, called ``corona'', is characterized by lower energy density and is governed by standard physics mechanisms in vacuum, including standard vacuum hadronization~\cite{Becattini:2008ya}.
The effectiveness of generators including 
a core-corona procedure with microcanonical hadronization of the core, like e.g. \texttt{EPOS4}, in describing the ALICE data has been assessed~\cite{Werner:2023jps}. 
 The core-corona model has been explored in EAS framework too~\cite{Baur:2019cpv}, using a modified version~\cite{Pierog:2023ahq} of the 
\texttt{EPOS-LHC} event generator. 
On the other hand, interplays between core and corona and a dynamical core-corona scenario have been explored in Ref.~\cite{Kanakubo:2021qcw, Kanakubo:2022jju}.   It has also been observed that SMC generators with more sophisticated string interaction and fragmentation mechanisms than \texttt{PYTHIA}, without including any QGP phase, like e.g. \texttt{DIPSY}~\cite{Gustafson:2018mhz}, can also go in the right direction by exploiting information on string position and overlap. 

A crucial investigation 
that will certainly help to understand the origin of collective phenomena in small systems, discriminating between QGP scenarios and non-QGP ones, concerns the experimental measurement of strange meson and baryon production at large rapidity. The observations of strangeness enhancement by the ALICE experiment are in fact limited to central production. On the other hand, measurements of the $x_F$ distribution of $\eta$ mesons by LHCf have enlightened a mismodeling of strangeness production in the very forward region by most of the hadronic-interaction generators used in EAS description, which would produce an excess of forward strangeness with respect to the measurement~\cite{Piparo:2023yam}. LHCb measurements of strange hadron production, covering rapidities $2 < y < 4.5$, would indeed be very welcome. This is especially important even for EAS physics, considering that no rapidity cuts affect the observation of particles produced by CR interactions with the air. 

Even the strangeness content of the proton and nuclei is a topic of intense debate~\cite{Accardi:2016ndt, Faura:2020oom}, which might be relevant especially when considering forward strangeness production. At the moment strange quark PDFs and nPDFs are accompanied by large uncertainties, having been constrained only in a limited way by the available experimental data (in particular, legacy data on charm production in $\nu$-induced deep-inelastic-scattering (DIS) processes and Drell-Yan production data in fixed-target experiments and at hadron colliders)~\cite{Alekhin:2017olj}, also considering some tension between datasets from different experiments/techniques. 
We expect that important constraints may come from forthcoming $\nu$-induced DIS experiments, exploiting e.g. the LHC beams of forward neutrinos produced in $pp$ scattering at the LHC. Two experiments, Faser$\nu$~\cite{FASER:2019dxq} and SND@LHC~\cite{SHiP:2020sos}, located at a $\sim$ 480 m distance from the ATLAS interaction point, on opposite sides, are already studying the interaction of these neutrinos with suitable detectors. Further enhanced versions of these experiments have been proposed for the HL-LHC phase, together with the possibility of building a new dedicated facility, the Forward Physics Facility~\cite{Anchordoqui:2021ghd}, ho\-sting multiple detectors.  
Their ability to distinguish $\nu$ and $\bar{\nu}$ will also be quite useful to assess the size of the $s(x)-\bar{s}(x)$ asymmetry.

On the other hand, these experiments offer the possibility to distinguish between $\nu$ and $\bar{\nu}$ of different flavours, 
measuring their forward fluxes as a function of $E_\nu$. This will allow to access  the charged $K/\pi$ ratio at forward rapidity, considering that, for $E_\nu < 200$ GeV, $\phi(\nu_e)/\phi(\nu_\mu)$ can be considered a proxy for $K^+/\pi^+$ ratios.

Besides $K/\pi$ ratios by themselves,
further measurements that can be very helpful to discriminate between different mechanisms for strangeness enhancement are 
the correlations of $K/\pi$ 
with the charged particle multiplicity $N_{ch}$. In this case the predictions from standard SMC event generators, the event generators including QGP elements and those including advanced mechanisms like color reconnection, string shoving and rope hadronization significatively differ among each other~\cite{Scaria:2023coa}.
On the other hand, inconsistencies have also been observed between charged kaon production data from different fixed-target experiments, e.g. NA61 investigating
$\pi^- C$ collisions and NA49 investigating $pp$ ones, considering the difficulty of e.g. the \texttt{QGSJET-III} event generator~\cite{Ostapchenko:2024jsg} to reproduce both datasets at a same time~\cite{Ostapchenko:2024myl}. 

Other interesting observables explorable at colliders are the correlations between the ratio of the electromagnetic and hadronic energy of events as a function of rapidity $R(\eta) = \langle dE_{em}/d\eta \rangle / \langle dE_{had}/d\eta \rangle$ and $N_{ch}$ or charged $K/\pi$ ratios~\cite{Scaria:2023coa}. These measurements can give very valuable insights on the performances of the event generators for hadronic interactions used for LHC and for EAS studies, and on the way to improve them. 

A third mechanism to reduce $f_{\pi^0}$ is enhancing light-baryon production, by replacing e.g. charge-neutral combinations of two or three pions, with $p\bar{p}$, $n\bar{n}$ systems.  
At this point one has to observe that data for production of $p$ and $\bar{p}$ 
are not reproduced well, or equally well, by event generators for hadronic interactions used in EAS simulations. \texttt{QGSJET-II} for instance is not able to reproduce neither the NA61 $\pi + C \rightarrow p/\bar{p}$ data at large momentum, nor the LEBC-EHS $\pi + p \rightarrow p/\bar{p}$ data at large $x_F$, whereas the recently released \texttt{QGSJET-III} is able to reproduce the LEBC-EHS $\bar{p}$ data, but not the
NA61 $\bar{p}$ ones, and is still not able to reproduce neither the LEBC-EHS nor the NA61 $p$ data. 
On the one hand, this might point to inconsistencies between datasets. On the other hand, it might be a signal of the need of different hadronization mechanisms. Nowadays there is increasing awareness, even in the collider community, that hadronization and its interplay to parton showers needs better modeling~\cite{Andersen:2024czj}, going beyond the string and cluster mechanisms currently implemented in generators. Attempts to go beyond the standard string hadronization mechanism by including e.g. color reconnection and string shoving effects have already been considered and would enhance the production of baryons, not necessarily carrying strangeness. On the other hand, string rope formation would enhance even strangeness. The development of parton shower algorithms in medium (going beyond what is currently done in vacuum) would also offer an interesting alternative for a microscopic description of $p$-ion and ion-ion collisions.

Antiproton production has also been investigated in the LHCb-SMOG apparatus, in $pHe$ collisions at $\sqrt{S_{NN}} = 110$~GeV, also measuring the ratio between the prompt and the detached production, with the latter occurring via hyperon 
decay~\cite{LHCb:2018ygc}. The data show strangeness enhancement with respect to both the Monte Carlo generators for hadronic interactions used at colliders and those used in EAS. Further measurements with different targets are foreseen. This is important not only for improving event generators but even for the interpretation of the quite precise measurements of antiproton fluxes performed by PAMELA and AMS02 at low energies.
 
\section{Forthcoming $p$O and OO runs at the LHC}

The LHC is also planning short $p$O and OO runs~\cite{Dembinski:2020dam}. This will allow the measurement of the degree of strangeness enhancement in inelastic collisions using a light target abundant in air, as well as measurements of distributions of pions, $\rho$'s, etc. Even the measurement of the total inelastic cross section for $p$-O collisions can be performed and it will be instructive to compare it to the results of the measurements in $p$-Air from the Pierre Auger Observatory and projections for $\sigma_{TOT, inel}(p$-Air) from measurements of  $\sigma_{TOT, inel}(pp)$ by forward experiments. In this respect, it is worth mentioning the discrepancy between the TOTEM measurements and the nore recent ATLAS ALFA ones~\cite{ATLAS:2022mgx}, lower by a few mb. The precise value of $\sigma_{TOT, inel}(p$-Air) affects the development of EAS and EAS observables.  
We expect all large experiments at the LHC to be active and take data during the Oxygen runs. On the other hand, the hypothesis of using different targets, such as Ne, although technically feasible at the LHC, so far has not been scheduled for any forthcoming run. Additionally, complementary measurements of the products of inelastic scattering of $p$ with O and a number of other gaseous targets will be achievable with the SMOG2 apparatum at LHCb, making use of just one LHC $p$ beam impinging on a fixed gaseous target (O can be used as alternative to many other gases). The collisions probed at LHCb-SMOG2 occur at $\sqrt{S_{NN}} \sim \mathcal{O}(100$ GeV), corresponding to intermediate stages of EAS generated by UHE CR, when secondaries interact with the atmosphere, with collision energy decreasing with increasing generation. 

\section{Diffraction}
The modelization of very forward scattering processes, including diffraction, in the event generators is also a big challenge.  In the past, no Monte Carlo generator was able to reproduce the energy distribution of far-forward neutrons
with $\eta > 10.75$ seen by the LHCf, but the agreement was qualitatively better for $\eta < 9$~\cite{LHCf:2020hjf}. It has then be understood that the issue was connected with the mismodeling (or complete lack of modelling) of diffractive processes with $\pi^0$ exchange. A recent study with \texttt{QGSJET-III} has shown that the agreement of forward neutron production with LHCf experimental data has
improved with respect to \texttt{QGSJET-II} thanks to the incorporation of this mechanism~\cite{Ostapchenko:2024myl}, although shape differences are still visible in the peak and/or tail of the energy distribution in rapidity bins $\eta > 9.65$. 
 Specialized forward experiments like the Zero Degree Calorimeters (ZDC) and the Forward Proton Spectrometers (FPS) allow to access single diffractive processes with $\pi^0$ exchange and pomeron exchange, measuring forward neutron and proton production, respectively. This will be particularly relevant during the runs with O discussed in the previous section.

\section{Insights on particle production in regions not covered at colliders by EAS-related observables}
In order to understand how to improve the generators for hadronic interactions, in the attempt to solve the muon puzzle, it has also been proposed to look at EAS observables related simultaneously to both $N_\mu$ and $\langle X_{max} \rangle$. An example is given in the following. 
Dividing the ($\ln N_\mu$, $\langle X_{max} \rangle$) plane in stripes, corresponding to $\langle X_{max} \rangle$ intervals, it is possible to look
at $\Lambda_\mu$, the slope of the $\ln N_\mu$ distribution at low $\ln N_\mu$
for each $\langle X_{max} \rangle$ bin~\cite{Cazon:2024cre}. 
 $\Lambda_\mu$ increases with $\langle X_{max} \rangle$, with differences among generators increasing too. 
Flatter
 low-$\ln N_\mu$ tails correspond to EAS characterized by less abundant, and then softer, hadronic activity.
Smaller $\langle X_{max} \rangle$ values correspond on average to EAS with larger hadronic activity and smaller electromagnetic activity. The opposite is true for larger $\langle X_{max} \rangle$ values. Given that $\Lambda_\mu(X_{max})$ can be measured, it can be used to tune the hadronic interaction models, in particular e.g. as for the energy spectrum of $\pi^0$ in regions hardly covered or not covered by detectors at human-made accelerators. This is a clear example where astrophysical measurements can offer a useful complement with respect to measurements at man-made accelerators, considering the limited kinematic range of the latter.

\section{Conclusions}
We expect many new important measurements that will allow on the one hand to better understand elementary particle physics, and, on the other one, to better 
constrain UHECR physics. 
The $\mu$ puzzle can be probably solved by considering a number of small effects.
Besides LHC measurements (with all possible detectors, even including those for
fixed-target collisions and those for far-forward $\nu$, and with a particular attention to the forthcoming runs with O), studies at the forthcoming 
electron-ion collider (EIC) will allow to investigate initial conditions for QGP formation, 
collectivity phenomena in small systems, as well as radiation and hadronization in the nuclear medium,
providing complementary precious information.
On the other hand, even new measurements in EAS experiments (e.g. measurements
of 2-dimensional distributions on a shower-by-shower basis, muon measurements 
 as a function of primary CR zenith angle, measurements of 
$\Lambda_\mu (X_{max})$, as well as muon measurements by different arrays 
will help to improve the generators for hadronic interactions
and hopefully clarify  the sources of the $\mu$ puzzle, with the consequence of an improved understanding of the CR composition and origin. This, in turn, will affect the interpretation of the data of all experiments where the interaction of CRs with the atmosphere produce relevant backgrounds. 

\bibliographystyle{JHEP}
\bibliography{crismac_proc_mvg_ar}

\end{document}